\documentclass{PoS}
\pdfoutput=1

\usepackage{amsmath}
\usepackage{here}
\usepackage{graphicx}
\usepackage{epsfig}
\usepackage{subcaption}

\title{Attacking One-loop Multi-leg Feynman Integrals with the Loop-Tree Duality}

\ShortTitle{Loop-Tree Duality}

\author{\speaker{Grigorios Chachamis}\\
        Instituto de F{\' \i}sica Te{\' o}rica UAM/CSIC, Nicol{\'a}s Cabrera 15\\
        \& Universidad Aut{\' o}noma de Madrid, E-28049 Madrid, Spain.\\
        E-mail: \email{chachamis@gmail.com}}
        
\author{Sebastian Buchta\\
        Instituto de F\'{\i}sica Corpuscular, Universitat de Val\`{e}ncia 
-- Consejo Superior de Investigaciones Cient\'{\i}ficas, 
Parc Cient\'{\i}fic, E-46980 Paterna, Valencia, Spain.\\
        E-mail: \email{sbuchta@ific.uv.es}}

\author{Petros Draggiotis\\
Institute of Nuclear and Particle Physics, NCSR "Demokritos", 
Agia Paraskevi, 15310, Greece.\\
E-mail: \email{petros.draggiotis@gmail.com}}

\author{Germ\'an Rodrigo\\
Instituto de F\'{\i}sica Corpuscular, Universitat de Val\`{e}ncia 
-- Consejo Superior de Investigaciones Cient\'{\i}ficas, 
Parc Cient\'{\i}fic, E-46980 Paterna, Valencia, Spain.\\
E-mail: \email{german.rodrigo@csic.es}}

\abstract{We discuss briefly the
 first numerical implementation of the Loop-Tree Duality (LTD) method.
 We apply the LTD method in order to calculate 
 ultraviolet and infrared finite multi-leg one-loop Feynman integrals. 
We attack scalar and tensor integrals with up to six legs (hexagons). The LTD method 
shows an excellent performance independently of the number of external legs.}

\FullConference{XXIV International Workshop on Deep-Inelastic Scattering and Related Subjects\\
		11-15 April, 2016\\
		DESY Hamburg, Germany}

\begin{document}

\section{Introduction}
The progress in calculating hadronic processes 
to higher orders of the perturbative expansion 
and in particular, calculating LHC observables 
to next-to-leading order (NLO) accuracy
was really impressive in recent years. Naturally, the complexity
for any observable to be computed at fixed order is related to the number of
initial and final states present in the process and the specific
order of the radiative corrections one takes into account.

Typically, in order to finally have a theoretical estimate that can be compared to experimental data,
a number of steps needs to be completed before, each with its own difficulties:
setting up the virtual and real parts of the scattering 
amplitude, integrating over any loop momenta, dealing with the cancelation
of the infrared divergencies~\cite{Catani:1996jh}, integrating over phase-space, etc.
One path to follow toward automated NLO calculations is the
purely numerical approach which has seen considerable development in the past years
~\cite{Soper:1998ye,Soper:1999xk,Soper:2001hu,Kramer:2002cd,Ferroglia:2002mz,Nagy:2003qn,Nagy:2006xy,Moretti:2008jj,Gong:2008ww,Kilian:2009wy,Becker:2010ng,Becker:2012aqa,Becker:2012nk,Seth:2016hmv}.
Computations based on a solid algorithmic setup at NLO are now standard, 
either purely numerical~\cite{Bevilacqua:2011xh,Cascioli:2011va,Cullen:2014yla} 
or a mixture of analytical and numerical techniques~\cite{Frixione:2008ym,Gleisberg:2008ta,Alwall:2014hca}. 
Substantial progress has also been made at higher orders~\cite{Passarino:2001wv,Anastasiou:2007qb,Becker:2012bi}.

The loop--tree duality (LTD) method~\cite{Catani:2008xa,Rodrigo:2008fp,Bierenbaum:2010cy,Binoth:2010nha,Bierenbaum:2012th,Bierenbaum:2013nja,Buchta:2014dfa,Buchta:2014fva,Buchta:2015vha,Sborlini:2015uia,Buchta:2015wna,
Hernandez-Pinto:2015ysa,Sborlini:2016gbr,Sborlini:2016fcj}
states that loop integrals and amplitudes with loop diagrams  with $n$ external legs
may be decomposed into 
a sum of tree-level-like graphs 
which still would need to be integrated over a measure that 
 resembles the usual $(n+1)-$body three-dimensional phase-space~\cite{Catani:2008xa,Rodrigo:2008fp}. 
This brings forward the intriguing possibility that loop- and tree-like radiative corrections 
can be treated on equal footing under a common integral 
using Monte Carlo integration through a convenient mapping of
the external momenta entering the virtual and real scattering 
amplitudes~\cite{Hernandez-Pinto:2015ysa,Sborlini:2016gbr}.

Here we focus on the use of the LDT framework in
computing one-loop  Feynman diagrams. 
We discuss briefly the principle behind the LTD method, we describe
 our numerical implementation~\cite{Buchta:2015wna} and we give 
 some details on the performance of our code
in computing both scalar and tensor multi-leg integrals.

\section{ Numerical Implementation of the LDT}

A generic one-loop scalar integral can be written in dimensional regularisation as 
\begin{equation}
L^{(1)}(p_1, p_2,\dots , p_N) = \int\limits_{\ell}\prod\limits_{i=1}^NG_F(q_i)~,
\label{eq-a}
 \end{equation}
where $G_F(q_i) = 1/(q_i^2-m_i^2+i0)$ are Feynman propagators, 
 $q_i$ are internal momenta depending on the loop momentum $\ell$
 and $\int_{\ell} = -i\int d^d\ell/(2 \pi)^d$ is used as a shorthand notation.
 To apply the LTD we integrate over the  energy coordinate 
 of the loop four-momentum by taking residues (Cauchy's theorem) after
 choosing an integration contour which encloses the poles with positive energy and negative imaginary part.
  This action decomposes
  the initial diagram from an integral with loop four-momentum to
  a sum of integrals over three-momentum:
  \begin{figure}[H]
  \centering
  \begin{subfigure}[h]{0.3\textwidth}
  \begin{equation}
L^{(1)}(p_1,p_2,\dots ,p_N)= \nonumber
\end{equation}
\end{subfigure}
\begin{subfigure}[h]{0.69\textwidth}
  \includegraphics[scale=1]{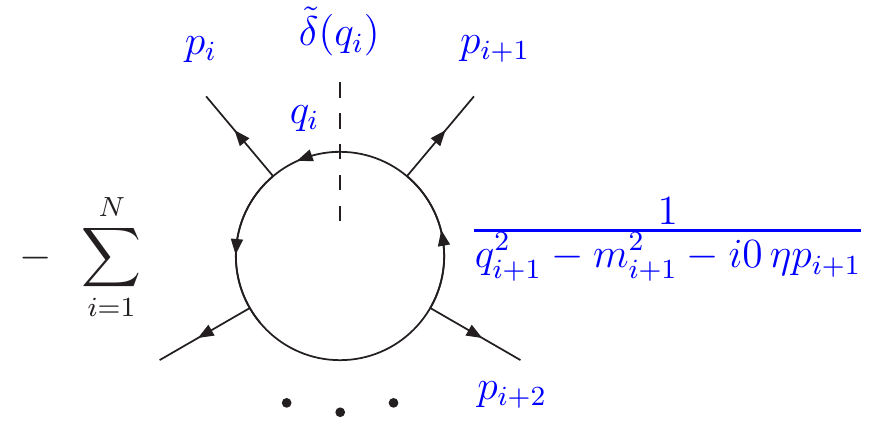}
    \end{subfigure}
  \end{figure}
  \hspace{-.7cm}We defined $\tilde{\delta}(q_i)=2\pi i\delta_+(q_i^2-m_i^2)$ where the 
  delta function  ``+''  subscript indicates that we take the positive-energy solution
  whereas, $\eta$ is a future-like vector. 
To summarise,  LTD in its essence tells us that by using Cauchy's theorem, me may rewrite a one-loop amplitude as a sum of single-cut phase-space integrals (which we call ``dual contributions'')
over the loop three-momentum. To integrate the dual contributions over the latter
requires a proper contour deformation due to the presence of the
so-called {\it ellipsoid} and {\it hyperboloid}  singularities~\cite{Buchta:2015wna}.

Our implementation of the LTD method was done in
\verb!C++!  and to perform the momenta integration
we employ the Cuba library~\cite{Hahn:2004fe}. 
The user needs only to specify the external four-momenta, the internal propagator masses and has
also the freedom to change the parameters of the contour deformation. 
One may also choose between Cuhre~\cite{Cuhre1, Cuhre2} and VEGAS~\cite{Lepage:1980dq}, 
and specify the desired number of evaluations. 
At run time, the code performs the following steps:
\begin{enumerate}
\item Reads in and assigns masses and external momenta.
\vspace{-.2cm}
\item Makes an analysis of the ellipsoid and hyperboloid singularities to
properly group the dual contributions and sets up the contour deformation.
\vspace{-.2cm}
\item Sends the integrand to the chosen routine (either Cuhre or VEGAS) for the
integration to be done and returns the result.
\end{enumerate}
\vspace{-.2cm}
To generate a very wide range of random momenta and masses 
so that we could test our code in various regions of the phase-space
we used {\tt MATHEMATICA 10.0}~\cite{Mathematica}.
For most of the numerical results in~\cite{Buchta:2015wna}, 
we used the routine Cuhre, VEGAS was a slower choice in general.
In order to produce reference values to compare our results to,
{\tt LoopTools 2.10}~\cite{Hahn:1998yk}
and  {\tt SecDec 3.0}~\cite{Borowka:2015mxa} 
were used.

We have run our code for a large number of scalar and 
up to rank three tensor integrals. We did an exhaustive comparison
of our results against the reference values from {\tt LoopTools 2.10}
and  {\tt SecDec 3.0} for triangles (3-external legs), boxes (4-external legs)
pentagons (5-external legs) and hexagons (6-external legs).
The execution time for scalar integrals with a wanted precision of 4-digits,
on a typical Desktop machine  (Intel i7 @ 3.4 GHz processor, 4-cores 8-threads),
varied from below a second to around 30 seconds, the latter for pentagons.
Tensor integrals were generally slower to compute but not by much.

As a non-trivial example, we display in Table~\ref{tab:hexagonlp} the results 
we obtained for tensor hexagons and for
three phase-space points (P22, P23, P24) along with the execution times.
The momenta  and the masses of the these phase-space points can be found
 in the Appendix of Ref.~\cite{Buchta:2015wna}.
The running times of SecDec are shown only for completeness, 
we do not imply that our code compares better or worse to SecDec.

\begin{table}[htb]
\begin{center}
\begin{tabular}{|ccllll|} \hline
& Rank & Tensor Hexagon & Real Part  &  Imaginary Part & Time[s]\\
\hline
P22
   & 1 & SecDec    & $~~1.01359(23) \times 10^{-15}$ & $+ i~2.68657(26) \times 10^{-15}$ & 33\\
   &   & LTD       & $~~1.01345(130)\times 10^{-15}$ & $+ i~2.68633(130)\times 10^{-15}$ & 72\\
\hline
P23 
   & 2 & SecDec    & $~~2.45315(24) \times 10^{-12}$ & $- i~2.06087(20) \times 10^{-12}$ & 337\\
   &   & LTD       & $~~2.45273(727)\times 10^{-12}$ & $- i~2.06202(727) \times 10^{-12}$ & 75\\
\hline
P24 
   & 3 & SecDec    & $-2.07531(19) \times 10^{-6}$ & $+ i~6.97158(56) \times 10^{-7}$ & 14280\\
   &   & LTD       & $-2.07526(8)  \times 10^{-6}$ & $+ i~6.97192(8) \times 10^{-7}$ & 85\\
\hline
\end{tabular}
\caption{Tensor hexagons involving numerators of rank one to three.
\label{tab:hexagonlp}}
\end{center}
\end{table}

\vspace{-1.cm}
\section{Conclusions}

The Loop-Tree Duality  is a method that incorporates 
many appealing theoretical features when considering processes with many external 
legs. The first numerical implementation
of the LTD  and its application to a very large number of phase-space test points
proved to be very 
successful.   
The code excels in diagrams
with many external legs as 
it shows only a moderate increase in execution times in comparison to diagrams
with smaller number of external particles. 

From this first study, and without any serious attempt 
toward the optimization of our code for faster execution times,
we conclude that our implementation of the 
LTD method offers a promising
alternative for computing multi-leg scalar and tensor one-loop integrals
with an arbitrary number of scales.
Finally, we are looking forward to testing our LTD numerical implementation
 onto physical processes.

\begin{flushleft}
\vspace{-.2cm}
{\bf \large Acknowledgements}
\end{flushleft}
\vspace{-.2cm}
We thank S. Catani for a longstanding fruitful collaboration. We also thank 
S. Borowka and G. Heinrich for their help in running SecDec. 
This work has been supported  
by the Spanish Government and ERDF funds from the European Commission (Grants
No. FPA2014-53631-C2-1-P, SEV-2014-0398, FPA2013-44773-P),
and by Generalitat Valenciana under Grant No. PROMETEOII/2013/007.
GC acknowledges support from the MICINN, Spain, 
under contract FPA2013-44773-P.
S.B. acknowledges support from JAEPre programme (CSIC). P.D. acknowledges support from 
General Secretariat for Research and Technology of Greece and from European Regional Development Fund 
MIS-448332-ORASY (NSRF 2007-13 ACTION, KRIPIS).

\end{document}